\documentclass[twocolumn,aps,epsfig]{revtex4}

\usepackage{graphicx}
\def\Slash#1{{\ooalign{\hfil$#1$\hfil\crcr\hfil$/$\hfil}}}

\begin{document}
\preprint{TU-620}
\title{Chiral Anomaly in Toroidal Carbon Nanotubes}
\author{K. Sasaki}
\affiliation{Department of Physics, Tohoku University, Sendai 980-8578, Japan}

\begin{abstract}
It is pointed out that the chiral anomaly in 
1+1 dimensions should be observed in toroidal
carbon nanotubes on a planar geometry with varying magnetic field. 
We show that the chiral anomaly is closely connected with the persistent 
current in a one-dimensional metallic ring.
\end{abstract}
\maketitle

Recently carbon nanotubes(CNTs)~\cite{Iijima}
have attracted much attention from various points of view.
Especially their unique mechanical and electrical properties have
stimulated many people's interest
in the analysis of CNTs~\cite{SDD,Dekker}. 
They have exceptional strength and stability,
and they exhibit either metallic or semiconducting behavior depending on
the diameter and helicity~\cite{MDW,Wildoer}.
Because of their small size, properties of CNTs should be 
governed by the law of quantum mechanics. 
Therefore it is quite important to understand the 
quantum behavior of electrons on CNTs. 
The bulk electric properties of (single-wall)CNTs are 
relatively simple,  
but the behavior of electrons at a metal-CNT junction is complicated
and its understanding is necessary for building actual 
electrical devices. On the other hands, toroidal carbon nanotubes
(Fullerence `Crop Circles'~\cite{Liu}, 
hereafter we use `torus' or `nanotorus' instead of 
`toroidal carbon nanotube' for simplicity) are clearly simple
because of their no-boundary shape and they would also exhibit either 
metallic or semiconducting properties.
Even in the torus case, quite important effect:
`chiral anomaly'~\cite{BJ}, which 
is of essentially quantum nature, might occur. 

Low energy excitations in CNTs at half filling 
move along the tubule axis because 
the circumference degree of freedom(an excitation in 
the compactified direction) is frozen by a wide energy gap. 
Hence this system can be described as a 1+1 dimensional system.
Furthermore in the case of metallic CNTs, the system describing
small fluctuations around the Fermi point is equivalent to two components 
``massless'' fermions in 1+1 dimensions. 
This situation can be modeled by the
quantum field theory of massless fermion which couples to 
the gauge field through minimal coupling.
This model realizes the chiral anomaly phenomenon~\cite{Schwinger,Manton,IM}.

The chiral anomaly is one of the most interesting phenomena in 
quantum field theory and has had an appreciable influence on the modern
development of high energy physics~\cite{Jackiw} and of condensed matter physics~\cite{NN}. 
The effect of the chiral anomaly on the electrons in a nanotorus appears
directly as a current flow.
On the other hand, 
it is known in solid state physics that a one-dimensional metallic ring shows 
the persistent current~\cite{Cheung,LC} in an appropriate experimental setting.
The current originating from the chiral anomaly shows the same magnetic field dependence 
to the persistent current. Therefore, the chiral anomaly in 1+1 dimensions 
is closely connected with the persistent current.

In this letter, we examine the anomaly effect on toroidal carbon nanotubes and
discuss how such an effect can be observed experimentally.

\begin{figure}
\includegraphics[scale=0.3]{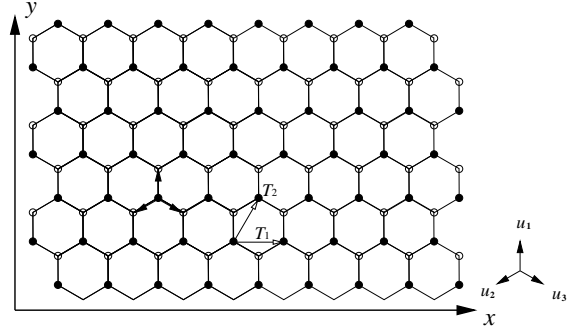}
\caption{Lattice structure of a two-dimensional graphite sheet
($ u_1 = a e_y,\ u_2 = -\frac{\sqrt{3}}{2}a e_x - \frac{1}{2}a e_y,\
u_3 = \frac{\sqrt{3}}{2}a e_x - \frac{1}{2}a e_y$)}
\label{fig:lattice}
\end{figure}

A CNT can be thought of as a layer of graphite sheet folded-up into a 
cylinder.
A graphite sheet consists of many hexagons whose vertices are occupied
by carbon atoms and each carbon supplies one conducting electron 
which determines the electric properties of the graphite sheet. 
The lattice structure of a two-dimensional graphite sheet is shown
in Fig.\ref{fig:lattice}. 
There are two symmetry translation vectors on 
this planar honeycomb lattice, $T_1 = \sqrt{3} a e_x ,
T_2 = \frac{\sqrt{3}}{2} a e_x + \frac{3}{2} a e_y$.
Here $a$ denotes the length of the nearest carbon vertex,
$e_x$ and $e_y $ are unit vectors which are orthogonal to each other 
($e_x \cdot e_y = 0$).
If we neglect the spin degrees of freedom,
because of these translation symmetries, the Hilbert space is spanned 
by the following two Bloch basis vectors,
\begin{eqnarray}
|\Psi^k_\bullet \rangle = \sum_{i\in \bullet} e^{ikr_i} a^\dagger_i |0 \rangle,
\ \
|\Psi^k_\circ \rangle = \sum_{i\in \circ} e^{ikr_i} a^\dagger_i |0 \rangle ,
\end{eqnarray}
where the black($\bullet$) and blank($\circ$) indices are 
indicated in Fig.\ref{fig:lattice}.
$r_i $ labels the vector pointing each site $i $, and 
$a_i,a_j^\dagger $ are canonically annihilation-creation operators of 
the electrons of site $i$ and $j$ that satisfy
$\{ a_i,a_j^\dagger \} = \delta_{ij}$.

We construct a state vector which is an eigenvector of these
symmetry translations as follows:
\begin{equation}
|\Psi^k \rangle = C^k_\bullet |\Psi^k_\bullet \rangle 
+ C^k_\circ |\Psi^k_\circ \rangle.
\end{equation}
In order to define the unit cell of wave vector $k$, 
we act the symmetry translation operators on the state vector and obtain
the Brillouin zone
\begin{equation}
-\frac{\pi}{\sqrt{3}} \le a k_x < \frac{\pi}{\sqrt{3}},\
-\pi \le \frac{\sqrt{3}}{2}a k_x + \frac{3}{2}a k_y < \pi ,
\end{equation}
where $k_x = k \cdot e_x $ and $k_y = k \cdot e_y $.
Now we compactify the sheet into a torus by imposing 
a boundary condition to the state vector.
For example, we may consider a zigzag type torus which has the 
following boundary conditions
\begin{eqnarray}
&&
\hat{G}(NT_1) |\Psi^k \rangle = |\Psi^k \rangle ,
\nonumber
\\
&&
\hat{G}(M(2T_2-T_1)) |\Psi^k \rangle = |\Psi^k \rangle.
\label{eq:bc}
\end{eqnarray}
$\hat{G} $ denotes a symmetry translation operator.
It is clear that there are many possibilities for the shape of the torus
and each shape has its own boundary condition. So,
some of them might have different properties from 
the above.
Especially we can image a torus in which some
twist exists along the tubule axis direction~\cite{CBJ}.
This system has the following boundary condition in general,
\begin{eqnarray}
\hat{G}(M(2T_2-T_1))|\Psi^k \rangle = 
\hat{G}(\tilde{N}T_1) |\Psi^k \rangle ,
\end{eqnarray}
where $\tilde{N} $ is determined by the 
twist at the junction of CNT ends.
Let us focus on the simple untwisted case given by Eq.(\ref{eq:bc}).
The periodic condition yields the discrete wave vectors
\begin{eqnarray}
a k_x = \frac{2\pi}{\sqrt{3}} \frac{n}{N}, \ \
a k_y = \frac{2\pi}{3} \frac{m}{M}.
\label{discretekx}
\end{eqnarray}
where $n$ and $m$ take an integer value.

Next we consider the Hamiltonian of this system~\cite{KM}.
Each carbon atom has an electron which makes $\pi $-orbital. 
The electron transfers from any site to the nearest 
three sites through the quantum mechanical 
tunneling or thermal hopping in finite temperature. 
Therefore there is some probability amplitude for this process.
In this case, the tight-binding Hamiltonian is most 
suitable.
\begin{equation}
{\cal H} = E_0 \sum_i a^\dagger_i a_i 
+ \gamma \sum_{\langle i,j \rangle} a_i^\dagger a_j,
\end{equation}
where the sum $\langle i,j \rangle $ is over pairs of 
nearest-neighbors carbon 
atoms $i,j$ on the lattice. $\gamma $ is the transition amplitude from 
one site to the nearest sites and $E_0 $ is the one from a site 
to the same site. The parameter $E_0 $ only fixes the origin of the 
energy and therefore is irrelevant. Hereafter we set $E_0 = 0$.

It is an easy task to find the energy eigenstates and eigenvalues 
of this Hamiltonian.
In the matrix representation, the energy eigenvalue equation reads
\begin{equation}
\pmatrix{ 0 & \gamma \sum_i e^{iku_i} \cr
\gamma \sum_i e^{-iku_i} & 0 \cr}
\pmatrix{ C_\bullet^k \cr C_\circ^k}
= E_k \pmatrix{ C_\bullet^k \cr C_\circ^k},
\end{equation}
where
$|\Psi^k_\bullet \rangle 
= (1,0)^t,\ |\Psi^k_\circ \rangle 
= (0,1)^t$,
and the vector $u_i $ is a triad of vectors pointing
respectively in the direction of the nearest neighbors of a 
black($\bullet$) cite shown in Fig.\ref{fig:lattice}. 
The energy eigenvalues and eigenvectors are as follows
\begin{eqnarray}
&&E_k = \pm \Delta(k), \\
&&\pmatrix{ C_\bullet^k \cr C_\circ^k}
= \frac{1}{\sqrt{2}\Delta(k)}
\pmatrix{ \gamma \sum_i e^{iku_i} \cr \pm \Delta(k)},
\end{eqnarray}
where
\begin{equation}
\Delta(k)= \gamma \sqrt{1+ 4 \cos{\frac{\sqrt{3}}{2}a k_x} 
\cos{\frac{3}{2}a k_y} + 4 \cos^2{\frac{\sqrt{3}}{2}a k_x}}.\nonumber
\end{equation}
The structure of this energy band has striking properties
when considered at half filling.
This is the situation which is physically interesting. Since 
each level of the band may accommodate two states due to the 
spin degeneracy, the Fermi level turns out to be at midpoint 
of the band ($E_k = 0 $).
Fermi points in the first Brillouin zone are located at $\tilde{k}_{1,2} =
(a k_x , a k_y) = ( \pm \frac{2\pi}{3\sqrt{3}} , \mp \frac{2\pi}{3})$.
Hence, if $N $ in Eq.(\ref{discretekx}) is a multiple of 3 then the torus shows 
metallic properties.

In order to understand the electric properties, we should take into
account a small perturbation around the Fermi point. 
So we take $k = \tilde{k}_1 + \delta k$
as a small fluctuation. Perturbation around the point $\tilde{k}_2 $ is 
same as around the point $k = \tilde{k}_1 $. 
So we may only consider one of the pairs.
In this case the effective Hamiltonian which describes the system is 
given by ${\cal H}_{pert} = v_F ( \sigma \cdot p ) $~\cite{note}
where $v_F (\equiv \frac{3\gamma a}{2\hbar})$ is the Fermi velocity, 
$p$ is the momentum operator ($p = -i \hbar \nabla $) and 
$\sigma_i$ are the Pauli matrices.
Hence the Schr\"odinger equation becomes
\begin{equation}
i\hbar \frac{\partial}{\partial t}\psi 
= v_F \left( \sigma \cdot p \right) \psi.
\end{equation}
We conclude that the low energy excitations of a metallic torus 
at half filling are described by an effective theory of two components
spinor obeying the Weyl equation. 

It should be noted that the characteristic properties of metallic 
CNTs are all reproduced quite well by analyzing this equation with 
external fields such as a magnetic and electronic field~\cite{DM}.
In the following, we consider metallic tori that have small $N$ and 
large $M$ values($M/N \sim 10^3 $).
In this case, transitions between different $ k_x$ are 
rarely happen because of 
their costed energy($\sim \gamma/N$) as compared to that 
of $k_y$:($\sim \gamma/M$).
Hence, the only surviving degree is a motion in the $y$-direction, 
i.e. this system is 1+1 dimensional effectively.

One can obtain the quantum field theory by promoting the wave 
function($\psi $) to the field operator($\Psi $) satisfying
the canonical anticommutation relations.
Because the Schr\"odinger equation is the Weyl equation 
it is appropriate to adopt the following Lagrangian density:
\begin{eqnarray}
{\cal L} = \bar{\Psi} \Slash{D} \Psi ,
\label{eq:Lagrangian}
\end{eqnarray}
where $\bar{\Psi} = \Psi^\dagger \gamma^0 $ and 
$\Slash{D}$ is the Feynman notation defined as 
$\Slash{D} = \sum_{\mu = 0,1} \left( i\hbar \partial_\mu - \frac{e}{c} A_\mu 
\right) \gamma^\mu $.
Here $A_\mu = (A_0,A_1) \equiv (A_t,v_F A_y) $ are the gauge fields and
we adopt the following relativistic notation:
$x^\mu = (x^0,x^1) \equiv (t,y/v_F), 
\partial_\mu = (\partial_0,\partial_1) \equiv \partial/\partial x^\mu$,
$\gamma^0 = \sigma_x,\ \gamma^1 = i\sigma_y,\ 
\gamma^5 =-\gamma^0 \gamma^1 = \sigma_z $.
The Dirac matrices $\gamma^\mu $ obey $\{\gamma^\mu,\gamma^\nu \} = 2 g^{\mu\nu} $
and $\gamma^\mu \gamma^5 = \epsilon^{\mu\nu} \gamma_\nu $ 
with the metric $g^{\mu\nu}={\rm diag}(1,-1) $ and the antisymmetric tensor $\epsilon^{\mu\nu} $,
$\epsilon^{01} = \epsilon_{01} = 1 $.
The electro-magnetic interaction is introduced according to the minimal coupling.
The gauge fields propagate in four dimensional space-time so that 
the Coulomb potential is given by the standard long-range interaction.
As a gauge fixing, we take the Weyl gauge $A_0 =0$; in this case,
the Hamiltonian of the fermion becomes
\begin{equation}
{\cal H}_F = \Psi^\dagger h_F \Psi = \Psi^\dagger
\pmatrix{ i\hbar \partial_1 -\frac{e}{c} A_1 & 0 \cr
                  0 & -( i\hbar \partial_1 -\frac{e}{c} A_1) } \Psi.
\end{equation}
We neglect the one-dimensional long-range Coulomb interaction~\cite{EG} and 
regard the gauge field as a classical field here.
Even in this case, it does not lose the nature of anomaly.
We list some main results of the Hamiltonian; 
a detailed description of this system can 
be found in References~\cite{Manton,IM}.
The energy eigenvectors are given 
by (Hereafter let us use $x$ instead of $y $ as a label of the 
coordinate of a tubule axis direction)
\begin{eqnarray}
&&
h_F \psi_n \pmatrix{ 1 \cr 0 } = \epsilon_n \psi_n \pmatrix{ 1 \cr 0 },\
h_F \psi_n \pmatrix{ 0 \cr 1 } = -\epsilon_n \psi_n \pmatrix{ 0 \cr 1 },
\nonumber
\\
&&
\ \ \ \psi_n(x) = \frac{1}{\sqrt{L}} e^{-i\frac{e}{\hbar c} 
\int_0^x A_1(x') dx' -i\frac{\epsilon_n}{\hbar v_F} x},
\end{eqnarray}
where $L$ is the circumferential length of a torus $L = 3aM $ and 
$\epsilon_n $ are the energy eigenvalues.
Because we take the periodic boundary condition,
the following energy spectrum appears
\begin{equation}
\epsilon_n = \frac{2\pi\hbar v_F}{L}
\left[  n
 - \frac{e}{2\pi \hbar c}\oint A_1 dx \right] .
\end{equation}

\begin{figure}
\includegraphics[scale=0.4]{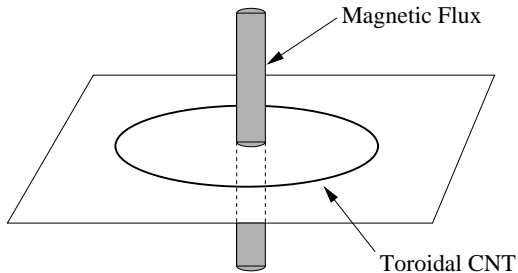}
\caption{A toroidal carbon nanotube on a planar geometry with a magnetic field}
\label{setup}
\end{figure}

The gauge field in the spectrum can be 
controlled externally by the following 
experimental setup. On the planar geometry 
we put a nanotorus
and penetrate some magnetic field inside the torus perpendicular
to the plane as is shown in Fig.\ref{setup}.
In this case the gauge field that expresses this magnetic field is given by,
in the vector notation,
$A = \frac{N_\Phi \phi_D}{2\pi} \nabla \theta$.
Therefore we get a component,
$A_1 = \frac{N_\Phi \phi_D}{L}$,
where $\phi_D = \frac{2\pi \hbar c}{e} $ is the flux quanta. 
This vector potential expresses $N_\Phi$ flux inside the torus and 
by tuning the magnetic field, $N_\Phi$ can be taken as a real number.

We expand the fermion field using the energy eigenvectors as
\begin{equation}
\Psi= \sum_{n\in Z} 
\left[ a_n \psi_n(x) e^{-i\frac{\epsilon_n}{\hbar}t}
\pmatrix{ 1 \cr 0} + b_n \psi_n(x) e^{+i\frac{\epsilon_n}{\hbar}t}
\pmatrix{ 0 \cr 1} \right], 
\end{equation}
where $a_n,b_n$ are independent fermionic annihilation operators 
satisfying the anti-commutators 
\begin{equation}
\{ a_n , a_m^\dagger \} = \{ b_n , b_m^\dagger \} = \delta_{nm}.
\end{equation}
All the other anticommutators vanish.
The dynamics of this field is governed by 
the Lagrangian density (\ref{eq:Lagrangian}), which 
has two conserved currents that are electric current $J^\mu $
and chiral current $J^\mu_5 $,
\begin{eqnarray}
&&
J^\mu(x) = \bar{\Psi}(x) \gamma^\mu \Psi(x),
\\
&&
J^\mu_5(x) = \bar{\Psi}(x) \gamma^\mu \gamma^5 \Psi(x) 
= \epsilon^{\mu\nu}J_\nu(x).
\label{eq:current}
\end{eqnarray}
Therefore, the following two charges conserve in the time evolution of the
system,
\begin{equation}
Q = \oint J^0(x)dx , 
\ \ Q_5 = \oint J^0_5(x)dx .
\end{equation}
Conservation of the electric current 
$\partial_\mu J^\mu=0$ ($\partial_0 = \partial_t,\partial_1 = v_F \partial_x $) 
is due to the gauge symmetry
and the chiral current conservation $\partial_\mu J_5^\mu=0$ is due to the global
chiral symmetry $\Psi \to e^{i\gamma^5 \alpha} \Psi$.
For unquantized fermion field the chiral invariance ensures conservation of the 
unquantized chiral current.
However after the second quantization the chiral current ceases to be conserved 
even though the interaction appears to be chirally invariant.
Because, different from classical mechanics,
in the world of quantum mechanics, the chiral symmetry is broken~\cite{BJ}
by the vacuum. So the chiral anomaly is similar to 
the spontaneous symmetry breaking 
in the sense that in both phenomena physical asymmetry is attributed to the 
vacuum state and not to the dynamics.

In order to find what is happening, we need to analyze the vacuum
structure $|vac;N_L,N_R \rangle = |vac;N_L \rangle \otimes |vac;N_R \rangle$,
where
\begin{equation}
|vac;N_L \rangle = \prod_{n=-\infty}^{N_L -1} a_n^\dagger |0\rangle,\ \ 
|vac;N_R \rangle = \prod^{n=\infty}_{N_R} b_n^\dagger |0\rangle.
\end{equation}
We define $|vac;N_L \rangle(|vac;N_R \rangle ) $ such that 
the levels with energy lower than $\epsilon_{N_L}(-\epsilon_{N_R -1})$ 
are filled and the others are empty.
On this vacuum, the expectation values of the charges 
and the energy become~\cite{Manton,IM}
\begin{eqnarray}
&&
\langle Q \rangle = N_L - N_R,
\label{eq:vev-Q}
\\
&&
\langle Q_5 \rangle = N_L + N_R -2N_\Phi -1,
\label{eq:vev-Q_5}
\\
&&
\langle H_F \rangle = \frac{2\pi \hbar v_F}{L}
\left( \frac{\langle Q \rangle^2 + \langle Q_5 \rangle^2}{4}
-\frac{1}{12} \right).
\label{eq:vev-H}
\end{eqnarray}

To obtain the above results,
we have regularized the divergent eigenvalues on the vacuum by 
$\zeta$-function regularization.
For example, the gauge charge is regularized as follows:
\begin{equation}
Q = \lim_{s\to 0} \left( \sum_{n \in Z} a_n^\dagger a_n 
\frac{1}{|\lambda \epsilon_n|^s}
+ \sum_{n \in Z} b_n^\dagger b_n 
\frac{1}{|-\lambda \epsilon_n|^s} \right),
\end{equation}
where $\lambda$ is an arbitrary constant with dimension 
of length which is necessary to make $\lambda \epsilon_n $ 
dimensionless. This regularization
respects gauge invariance because the energy of each level is a gauge 
invariant quantity. 

The gauge charge $\langle Q \rangle$ remains a constant 
if no electron flows into the system. 
We now have $ N_L = N_R $ for an isolated nanotorus.
From the above equation (\ref{eq:vev-Q_5}), 
it can be seen that if $N_L$ and $N_R$ are conserved, then,
by varying the magnetic field $N_\Phi$, the chiral charge also changes.
Therefore it is not a conserved quantity. 
We thus see that the vacuum is responsible for non-conservation of
chirality even though the dynamics is chirally invariant.
From Eq.(\ref{eq:current}) we see that 
the chiral current $J^0_5 $ is proportional to the electric current($e v_F J^1(x) $)
in the tubule axis direction, then we have an average value of the electric current $J$ as
\begin{equation}
J \equiv \frac{e v_F}{L}\oint  J^1(x)dx 
= - \frac{e v_F}{L} \oint J^0_5(x) dx = -\frac{e v_F}{L} Q_5.
\end{equation}
Hence, in order to observe the anomaly, we should observe the electrical 
current in the torus.

\begin{figure}
\vspace{0.5cm}
\includegraphics[scale=0.5]{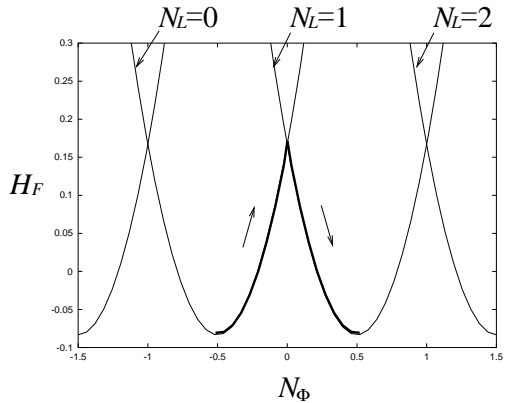}
\caption{Flux dependence of the fermionic vacuum energy $\langle H_F \rangle $.
The energy value is labeled in the unit of $\frac{2\pi \hbar v_F}{L}$.}
\label{fig:energy}
\end{figure}

\begin{figure}
\vspace{0.5cm}
\includegraphics[scale=0.5]{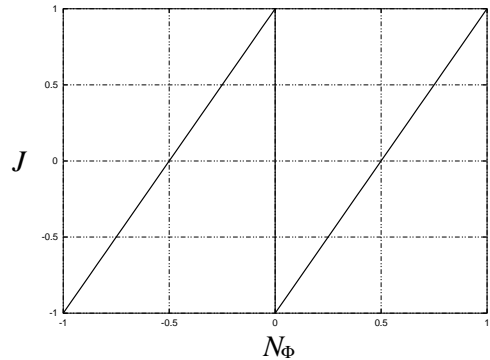}
\caption{Magnetic field dependence of the induced current. 
We plot the current in the unit of $\frac{ev_F}{L} $.}
\label{fig:current}
\end{figure}

It is clear from the above equations that there are two origins of the 
usual current flow along the torus. One is the $N_L + N_R$ term 
which can be induced in thermal bath or by a sudden change of the magnetic
field. On the other hand, the magnetic field can change the 
quantum vacuum structure and lead to the anomaly.
In order to avoid the unexpected 
changes of $N_L(=N_R) $, the magnetic field must 
be changed adiabatically at low temperature($< \frac{2\pi \hbar v_F}{L} $).
However, in an adiabatic process, when the strength of the
magnetic field reaches the point that $N_\Phi$ is an integer, then $N_L(=N_R)$
also have to change. The reason is that, when increasing $N_\Phi$ starting 
from the point $N_\Phi =-\frac{1}{2},N_L=0$, 
the energy is going up as Eq(\ref{eq:vev-H}).
At $N_\Phi = 0$, the spectrum meets an another line of spectrum 
coming from the $N_\Phi =\frac{1}{2},N_L=1$ as is shown in Fig.\ref{fig:energy}.
Therefore the circular current in the ring 
\begin{equation}
J = \frac{ev_F}{L} \left[ 2 \left( N_\Phi-N_L \right)  +1 \right]
\label{eq:J}
\end{equation}
follows the line shown in Fig.\ref{fig:current}.
We should remark that 
there are two spin degrees of freedom at each Fermi point.
Therefore the actual current is four times the $J$, that is,
the amplitude of this total current is $\frac{4ev_F}{L}$.
A numerical value of this amplitude is about $0.5 [{\rm \mu A}] $ for a nanotorus
with $L = 1[{\rm \mu m}] $.
This current for an untwisted torus shows the same magnetic field dependence to 
the persistent current in ref.~\cite{LC}.
Our results (\ref{eq:J}) are in agreement with the results of other papers.

Let us explain how to measure the current briefly.
Some methods could be considered in order to detect the current in the torus.
As an example, the current generates a magnetic field around torus, 
then one can observe the current via magnetic field which is 
generated by the current.  
However the current could not be observed by the standard electrical contact
because the electrical perturbation can not affect the current flow. 
This means that we can not measure the current by an electrical contact.

In conclusion,
low energy excitations in metallic toroidal carbon nanotubes
can be described by the two components
``massless'' fermion which couples to 
a gauge field through minimal coupling.
The anomaly effect should be observed by an 
adiabatic change of the vector potential, since 
this induces peculiar electrical current along the torus 
through the chiral anomaly. The chiral anomaly provides
a deeper understanding for the persistent current.

The author would like to thank M. Hashimoto and K. Hasebe
for fruitful discussion.
This work is supported by a fellowship of the Japan 
Society of the Promotion of Science.



\begin{thebibliography}{99}
\bibitem{note}
It should be noted that to obtain this Hamiltonian we have to 
perturb the Hamiltonian around the points $a\tilde{k}_1 - aK_1$
and $a\tilde{k}_2 + aK_1$. The wave vectors $K_i$ satisfy the conditions,
$ K_i \cdot T_j = 2\pi \delta_{ij}$.
The vectors 
$a\tilde{k}\pm aK_1$ and $a\tilde{k}$ represent the same state since any 
two wave vectors congruent by $aK_i$ are just different labels of the 
same state.
\end{thebibliography}
\end{document}